\documentstyle[aps,multicol,epsfig]{revtex}

\def\<{\langle}
\def\>{\rangle}
\def\tr{\mbox{Tr}}

\newcommand{\be}{\begin{equation}}
\newcommand{\ee}{\end{equation}}
\newcommand{\bea}{\begin{eqnarray}}
\newcommand{\eea}{\end{eqnarray}}

\title{Robustness of adiabatic quantum computation}
\author{Andrew M. Childs,$^{1}$\thanks{amchilds@mit.edu}
        Edward Farhi,$^{1}$\thanks{farhi@mit.edu}
	and John Preskill$^{2}$\thanks{preskill@theory.caltech.edu}}
\address{$^{1}$Center for Theoretical Physics, Massachusetts Institute of
               Technology, Cambridge, MA 02139, USA \\
	 $^{2}$Institute for Quantum Information, California Institute of
               Technology, Pasadena, CA 91125, USA}
\date{9 August 2001}

\begin{document}
\maketitle


\begin{abstract}
We study the fault tolerance of quantum computation by adiabatic
evolution, a quantum algorithm for solving various combinatorial search
problems.  We describe an inherent robustness of adiabatic computation
against two kinds of errors, unitary control errors and decoherence, and
we study this robustness using numerical simulations of the algorithm.

\hfill\mbox{[MIT-CTP \#3174; CALT-68-2324]}
\end{abstract}

\begin{multicols}{2}[]

\section{Introduction}

The method of quantum computation by adiabatic evolution has been proposed
as a general way of solving combinatorial search problems on a quantum
computer~\cite{farhi1}.  Whereas a conventional quantum algorithm is
implemented as a sequence of discrete unitary transformations that form a
quantum circuit involving many energy levels of the computer, the
adiabatic algorithm works by keeping the state of the quantum computer
close to the instantaneous ground state of a Hamiltonian that varies
continuously in time.  Therefore, an imperfect quantum computer
implementing a conventional quantum algorithm might experience different
sorts of errors than an imperfect adiabatic quantum computer.  In fact, we
claim that an adiabatic quantum computer has an inherent robustness
against errors that might enhance the usefulness of the adiabatic
approach.

The adiabatic algorithm works by applying a time-dependent Hamiltonian
that interpolates smoothly from an initial Hamiltonian whose ground state
is easily prepared to a final Hamiltonian whose ground state encodes the
solution to the problem.  If the Hamiltonian varies sufficiently slowly,
then the quantum adiabatic theorem guarantees that the final state of the
quantum computer will be close to the ground state of the final
Hamiltonian, so a measurement of the final state will yield a solution of
the problem with high probability.  This method will surely succeed if the
Hamiltonian changes slowly.  But how slow is slow enough?

Unfortunately, this question has proved difficult to analyze in general.
Some numerical evidence suggests the possibility that the adiabatic
algorithm might efficiently solve computationally interesting instances of
hard combinatorial search problems, outperforming classical
methods~\cite{farhi1,farhi2,farhi3,farhi4}.  Whether the adiabatic
algorithm provides a definite speedup over classical methods remains an
interesting open question.  As we will discuss in Sec.~\ref{sec:adi_qc},
the time required by the algorithm for a particular instance can be
related to the minimum gap $\Delta$ between the instantaneous ground state
and the rest of the spectrum.  Roughly speaking, the required time goes
like $\Delta^{-2}$.  Thus, if $\Delta^{-2}$ increases only polynomially
with the size of the problem, then so does the time required to run the
algorithm.  However, determining $\Delta$ has not been possible in
general.

Our objective in this paper is not to explore the computational power of
the adiabatic model, but rather to investigate its intrinsic {\em fault
tolerance}.  Since quantum computers are far more susceptible to making
errors than classical digital computers, fault tolerant protocols will be
necessary for the operation of large-scale quantum computers.  General
procedures have been developed that allow any quantum algorithm to be
implemented fault tolerantly on a universal quantum computer~\cite{fault},
but these involve a substantial computational overhead.  Therefore, it
would be highly advantageous to weave fault tolerance into the design of
our quantum hardware.

We therefore will regard adiabatic quantum computation not as a convenient
language for describing a class of quantum circuits, but as a proposed
physical implementation of quantum information processing.  We do not cast
the algorithm into the conventional quantum computing paradigm by
approximating it as a sequence of discrete unitary transformations acting
on a few qubits at a time.  Instead, suppose we can design a physical device
that implements the required time-dependent Hamiltonian with reasonable
accuracy.  We then imagine implementing the algorithm by slowly changing
the parameters that control the physical Hamiltonian.  How well does such
a quantum computer resist decoherence, and how well does it perform if the
algorithm is imperfectly implemented?

Regarding resistance to decoherence, we can make a few simple
observations.  The phase of the ground state has no effect on the efficacy
of the algorithm, and therefore dephasing in the energy eigenstate basis
is presumably harmless.  Only the interactions with the environment that
induce transitions between eigenstates of the Hamiltonian might cause
trouble.  In principle, these can be well controlled by running the
algorithm at a temperature that is small compared to the minimum gap
${\Delta}$.  (We use units in which Boltzmann's constant $k_B=1$, so that
temperature has units of energy.)  If $\Delta$ decreases slowly as the
size of the problem increases, then the resources required to run at a
sufficiently low temperature may be reasonable.  Since the adiabatic
method is only efficient if $\Delta$ is not too small, we conclude that
whenever the method works on a perfectly functioning quantum computer, it
is robust against decoherence.

In addition to environmental decoherence, we must also consider the
consequences of imperfect implementation.  Our chosen algorithm may call
for the time-dependent Hamiltonian $H(t)$, but when we run the algorithm,
the actual Hamiltonian will be $H(t)+K(t)$, where $K(t)$ is an ``error.''
An interesting feature of adiabatic quantum computation is that $K(t)$
need not remain small during the evolution in order for the algorithm to
work effectively.  A reasonably large excursion away from the intended
Hamiltonian is acceptable, as long as $K(t)$ is slowly varying and has
initial and final values that are not too large.  A very rapidly
fluctuating $K(t)$ may also be acceptable, if the characteristic frequency
of the fluctuations is large compared to the energy scale of $H(t)$. 

In this paper, we use numerical simulations to investigate the sensitivity
of an adiabatic computer to decohering transitions and to a certain class
of unitary perturbations induced by a Hamiltonian $K(t)$.  The results are
consistent with the idea that the algorithm remains robust as long as the
temperature of the environment is not too high and $K(t)$ varies either
sufficiently slowly or sufficiently rapidly.  Thus, the adiabatic model
illustrates the principle that when the characteristics of the noise are
reasonably well understood, it may be possible to design suitable quantum
hardware that effectively resists the noise.  However, note that some of
the effects of decoherence and unitary control error may not be
significant for the small problems we are able to study --- especially in
the case of decoherence, where the time required by the simulation
restricts us to systems with only four qubits --- and hence our data may
not be indicative of the performance of the algorithm working on larger
inputs.

A technique closely related to adiabatic computation was described by
Kadowaki and Nishimori~\cite{KN98} and has been tested experimentally (in
conjunction with a cooling procedure) by Brooke et al.~\cite{BBRA99}.  In
a different guise, the principles that make quantum adiabatic evolution
robust also underlie the proposal by Kitaev~\cite{kitaev} to employ
non-abelian anyons for fault-tolerant quantum computation.  The fact that
adiabatic evolution incorporates a kind of intrinsic fault tolerance has
also been noted
in~\cite{preskill_ft,ogburn,zanardi,ekert,lloyd_ft,freedman}.

In Sec.~\ref{sec:adi_qc} we review the adiabatic model of quantum
computation, and in Sec.~\ref{sec:ec3} we describe the specific
combinatorial search problem (three-bit exact cover) that we use in our
simulations.  Secs.~\ref{sec:decoherence} and~\ref{sec:control_error}
report our numerical results on decoherence and unitary control error, and
Sec.~\ref{sec:conclusions} summarizes our conclusions.

\section{Adiabatic quantum computation}
\label{sec:adi_qc}

We briefly review the adiabatic model of quantum computation introduced
in~\cite{farhi1}.  Let $h(z)$ be a function of $n$ bits $z=(z_1,z_2,z_3,
\dots, z_n)$, and consider the computational problem of finding a value of
$z$ that minimizes $h(z)$.  We will typically be interested in the case
were this value of $z$ is unique.  We may associate with this function the
Hermitian operator
\be
\label{ham}
  H_P=\sum_{z=0}^{2^n-1} h(z) |z\>\<z|
\,,
\ee
so that the computational basis state $|z\>$ is an eigenstate of $H_P$
with eigenvalue $h(z)$. Then the problem is to determine which state
$|z\>$ is the ground state (eigenstate with lowest eigenvalue) of $H_P$.
We refer to $H_P$ as the {\em problem Hamiltonian}.

The strategy for finding the ground state of $H_P$ is to prepare the
ground state of some other {\em beginning Hamiltonian} $H_B$ and slowly
interpolate to $H_P$.  A simple choice for the interpolation is given by
the one-parameter family of Hamiltonians
\be
  \tilde H(s) = (1-s) H_B + s H_P
\label{eq:full_ham}
\ee
that interpolates between $H_B$ and $H_P$ as $s$ varies from 0 to 1.  We
prepare the ground state of $H_B$ at time $t=0$, and then the state
evolves from $t=0$ to $t=T$ according to the Schr\"odinger equation,
\begin{equation}
  i{{\mathrm d} \over {\mathrm d}t}|\psi(t)\> = H(t)|\psi(t)\>
\,,
\end{equation}
where the Hamiltonian is
\begin{equation}
  H(t)= \tilde H(t/T)
\,.
\label{eq:timedep}
\end{equation}
At time $T$ (the {\em run time} of the algorithm), we measure the state in
the computational basis.  If we let $|\varphi\>$ denote the (unique)
ground state of $H_P$ for a given instance of the problem, then the {\em
success probability} of the algorithm for this instance is
\begin{equation}
  {\rm Prob}(T) \equiv |\<\varphi|\psi(T)\>|^2
\,.
\end{equation}

Does the algorithm work?  According to the quantum adiabatic
theorem~\cite{kato,messiah}, if there is a nonzero gap between the ground
state and the first excited state of $\tilde H(s)$ for all $s\in[0,1]$,
then ${\rm Prob}(T)$ approaches 1 in the limit $T\to\infty$.  Furthermore,
level crossings are non-generic in the absence of symmetries, so a
non-vanishing gap is expected if $H_B$ does not commute with $H_P$.  Thus,
the success probability ${\mathrm Prob}(T)$ of the algorithm will be high
if the evolution time $T$ is large enough.  The question is: how large a
$T$ is large enough so that ${\mathrm Prob}(T)$ is larger than some fixed
constant?

We can reformulate this question in terms of
\begin{equation}
  \Delta = \min_{s\in [0,1]} (E_1(s) - E_0(s))
\end{equation}
and
\begin{equation}
  {\cal E} = \max_{s\in [0,1]} \left| \<1,s|
                               {{\mathrm d}\tilde H\over {\mathrm d}s }
                               |0,s\> \right|
\,,
\end{equation}
where $E_0(s)$ is the lowest eigenvalue of $\tilde H(s)$, $E_1(s)$ is the
second lowest eigenvalue, and $|0,s\>$, $|1,s\>$ are the corresponding
eigenstates.  By calculating the transition probability to lowest order in
the adiabatic expansion~\cite{messiah}, one finds that the probability of
a transition from ground state to first excited state is small provided
that the run time $T$ satisfies
\begin{equation}
  T \gg {{\cal E} \over \Delta^2}
\,.
\label{eq:adiab_cond}
\end{equation}
If the spectrum consists of only two levels, then this condition is
sufficient to ensure that the system remains in the ground state with high
probability.  In general, the required run time $T$ will be bounded by a
polynomial in $n$ so long as $\Delta$ and $\cal E$ are polynomially
bounded.  For the problems we are interested in, $\cal E$ is polynomially
bounded, so we only have to consider the behavior of $\Delta$.

By rescaling the time, we can think of the evolution as taking place in
the unit time interval between $s=0$ and $s=1$, but in that case the
energy eigenvalues are rescaled by the factor $T$.  Roughly speaking, we
can think of ${\mathrm d} \tilde H(s)/{\mathrm d}s$ as a perturbation that
couples the levels of the instantaneous Hamiltonian $\tilde H(s)$, and has
the potential to drive a transition from $|0,s\rangle$ to $|1,s\rangle$.
But if $T$ is large, the effects of this perturbation are washed out by
the rapid oscillations of the relative phase $\exp[-i T\int_0^s {\mathrm
d}s' (E_1(s')-E_0(s'))]$.

Note that the Hamiltonian can be regarded as reasonable only if it is
``local,'' that is, if it can be expressed as a sum of terms, where each
term acts on a bounded number of qubits (a number that does not grow with
$n$).  Indeed, in this case, the Hamiltonian evolution can be accurately
and efficiently simulated by a universal quantum computer~\cite{lloyd}.
Many combinatorial search problems (e.g., 3SAT) can be formulated as a
search for a minimum of a function that is local in this sense.  Along
with a local choice of $H_B$, this results in a full $H(t)$ that is also
local.

A direct physical implementation of the continuously varying $H(t)$ would
presumably be possible only under a somewhat stronger locality condition.
We might require that each qubit is coupled to only a few other qubits, or
perhaps that the qubits can be physically arranged in such a way that the
interactions are spatially local.  Fortunately, there are interesting
computational problems that have such forms, such as 3SAT restricted to
having each bit involved in only three clauses or the problem of finding
the ground state of a spin glass on a cubic lattice~\cite{local}.
However, for the purposes of our simulation, we will only consider small
instances, and since we do not have a specific physical implementation in
mind, we will not concern ourselves with the spatial arrangement of the
qubits.

\section{An example: The exact cover problem}
\label{sec:ec3}

For definiteness, we study the robustness of the adiabatic algorithm via
its performance on the problem known as ``three-bit exact cover'' (EC3).
An $n$-bit instance of EC3 consists of a set of clauses, each of which
specifies three of the $n$ bits.  A clause is said to be satisfied if and
only if exactly one of its bits has the value 1.  The problem is to
determine if any of the $2^n$ assignments of the $n$ bits satisfies all of
the clauses.

For this problem, the function $h(z)$ is a sum
\begin{equation}
  h(z)=\sum_C h_C(z_{i_C},z_{j_C},z_{k_C})
\end{equation}
of three-bit clauses, where
\begin{eqnarray}
  && h_C(z_{i_C},z_{j_C},z_{k_C})\nonumber\\ 
  && \quad=\cases{0~,\quad (z_{i_C},z_{j_C},z_{k_C})
                  {\mathrm ~satisfies~clause~} C \cr
	          1~,\quad (z_{i_C},z_{j_C},z_{k_C})
	          {\mathrm ~violates~clause~} C \,.}
\end{eqnarray}
The value of the function $h(z)$ is the number of clauses that are
violated; in particular, $h(z)=0$ if and only if $z$ is an assignment that
satisfies all the clauses.

To solve EC3 by the adiabatic algorithm, a sensible choice for the
beginning Hamiltonian is
\begin{equation}
  H_B=\sum_C H_{B,C}
\,,
\label{eq:hb}
\end{equation}
where
\bea
  H_{B,C} = && {1\over 2} \left(1-\sigma_x^{(i_C)}\right)
              +{1\over 2} \left(1-\sigma_x^{(j_C)}\right) \nonumber\\
	    &&+{1\over 2} \left(1-\sigma_x^{(k_C)}\right)
\,,
\eea
which has the ground state
\begin{equation}
  |\psi(0)\> = {1\over 2^{n/2}}\sum_{z=0}^{2^n-1} |z\>
\,.
\label{eq:start_state}
\end{equation}
The resulting $H(t)$ is local in the sense that it is a sum of terms, each
of which acts on only a few qubits.  A stronger kind of locality can be
imposed by restricting the instances so that each bit is involved in at
most a fixed number of clauses.  The computational complexity of the
problem is unchanged by this restriction.

\begin{figure}
\begin{center}
\psfig{file=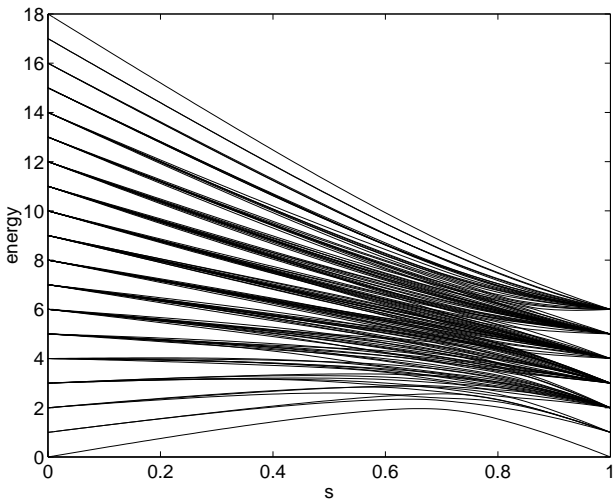,width=3.25in}
\end{center}
\caption{Spectrum of a randomly generated $n=7$ bit instance of EC3 with a
unique satisfying assignment.  Note that the energy gap between the ground
state and the first excited state is significantly larger than all other
gaps.  An expanded view would show that there are no level crossings
anywhere in the spectrum (except for the degeneracies at $s=0$ and
$s=1$).}
\label{fig:spectrum}
\end{figure}

Numerical studies of the adiabatic algorithm applied to this problem were
reported in~\cite{farhi2,farhi4}.  Instances of EC3 with $n$ bits were
generated by adding random clauses until there was a unique satisfying
assignment, giving a distribution of instances that one might expect to be
computationally difficult to solve.  The results for a small number of
bits ($n \le 20$) were consistent with the possibility that the adiabatic
algorithm requires a time that grows only as a polynomial in $n$ for
typical instances drawn from this distribution.  If this is the case, then
the gap $\Delta$ does not shrink exponentially.  Although the typical
spacing between levels must be exponentially small, since there are an
exponential number of levels in a polynomial range of energies, it is
possible that the gap at the bottom is larger.  For example,
Fig.~\ref{fig:spectrum} shows the spectrum of a randomly generated
seven-bit instance of EC3.  The gap at the bottom of the spectrum is
reasonably large compared to the typical spacing.  This feature is not
specific to this one instance, but is characteristic of randomly generated
instances, at least for $n \lesssim 10$, beyond which the repeated matrix
diagonalization required to create a picture of the spectrum becomes
computationally costly.  A large gap makes an instance readily solvable by
the adiabatic algorithm, and also provides robustness against thermal
transitions out of the ground state.

\section{Decoherence}
\label{sec:decoherence}

Perhaps the most significant impediment to building a large-scale quantum
computer is the problem of decoherence.  No quantum device can be
perfectly isolated from its environment, and interactions between a device
and its environment will inevitably introduce noise.  Fortunately, such
effects can be countered using fault-tolerant protocols, but as we have
already mentioned, these protocols can be costly.  Therefore, we would
like to consider quantum systems with inherent resistance to decohering
effects.  If the ground state of our adiabatic quantum computer is
separated from the excited states by a sizable energy gap, then we expect
it to exhibit such robustness.  Here, we consider how the adiabatic
algorithm for EC3 is affected by decoherence.

First, we briefly review the master equation formalism for describing the
decohering effects of an environment on a quantum system.  Suppose that
our quantum computer is a collection of spin-$1 \over 2$ particles
interacting with each other according to the Hamiltonian $H_S$ and weakly
coupled to a large bath of photons.  The total Hamiltonian of the quantum
computer and its environment is
\be
  H = H_S + H_E + \lambda V
\,,
\ee
where $H_E$ is the Hamiltonian of its environment, $V$ is an interaction
that couples the quantum computer and the photon bath, and $\lambda$ is a
coupling constant.  We may describe the state of the quantum computer
alone by the density matrix $\rho$ found by tracing over the environmental
degrees of freedom.  In general, the time evolution of $\rho$ is
complicated, but under reasonable assumptions, we can approximate its
evolution using a Markovian master equation.

One way of deriving such a master equation is to consider the weak
coupling limit, in which $\lambda \ll 1$~\cite{Dav74}.  If the environment
is very large and only weakly coupled to the quantum computer, it will be
essentially unchanged by the interaction.  Furthermore, in this limit, we
can expect the evolution of the quantum computer to be Markovian, or local
in time, if we filter out high-frequency fluctuations by some
coarse-graining procedure.  Assuming that the combined state of the
quantum computer and its environment begins in a product state $\rho(0)
\otimes \rho_E$, Davies derives the master equation
\be
  {{\mathrm d}\rho \over {\mathrm d}t} = -i[H_S,\rho] 
      + \lambda^2 K^\natural \rho
\,,
\label{eq:daviesmaster}
\ee
where
\bea
  K \rho &=& -\int_0^\infty {\mathrm d}x \, 
    \tr_E[U(-x) V U(x), [V, \rho \otimes \rho_E]] \\
  K^\natural \rho &=& \lim_{x \to \infty} {1 \over x} 
    \int_0^x {\mathrm d}y \, U(-y) \{ K [U(y) \rho U(-y)] \} U(y)
\label{eq:natural}
\eea
with
\be
  U(x) = e^{-i x (H_S+H_E)}
\,,
\label{eq:unitary}
\ee
where we have (temporarily) assumed that $H_S$ is time-independent.
Although the $\natural$ operation defined by (\ref{eq:natural}) does not
appear in some formulations of the Markovian master equation, it appears
to be essential for the equation to properly describe the weak coupling
limit~\cite{DS79}, and in particular, for it to capture the physics of
relaxation to thermal equilibrium.  The master equation
(\ref{eq:daviesmaster}) has the property that if the environment is in
thermal equilibrium at a given temperature, then the decohering
transitions drive the quantum computer towards the Gibbs state of $H_S$ at
that temperature.  While not an exact description of the dynamics,
(\ref{eq:daviesmaster}) should provide a reasonable caricature of a
quantum computer in a thermal environment.

Note that (\ref{eq:daviesmaster}) is derived assuming a time-independent
Hamiltonian $H_S$; with a time-varying $H_S(t)$, we should expect the
generator of time evolution at any particular time to depend on the
Hamiltonian at all previous times~\cite{DS78}.  However, if $H_S(t)$ is
slowly varying, then it is a good approximation to imagine that the
generator at any particular time depends only on $H_S$ at that
time~\cite{Lin83}.  In particular, since we are interested in nearly
adiabatic evolution, $H_S(t)$ varies slowly, so (\ref{eq:daviesmaster})
remains a good approximation, where at any given time $t$ we compute
$K^\natural$ using only $H_S(t)$.  Note that with $H_S(t)$ time-dependent,
$U(x)$ defined by (\ref{eq:unitary}) is not the time evolution operator;
it depends on the time $t$ only implicitly through $H_S(t)$.

For a system of spins coupled to photons, we choose the interaction
\be
  V = \sum_i \int_0^\infty {\mathrm d}\omega \, 
      [ g(\omega) a_\omega \sigma_+^{(i)} 
      + g^*(\omega) a_\omega^\dag \sigma_-^{(i)}]
\,,
\ee
where $\sum_i$ is a sum over the spins, $\sigma_\pm^{(i)}$ are raising and
lowering operators for the $i$th spin, $a_\omega$ is the annihilation
operator for the photon mode with frequency $\omega$, and $\lambda
g(\omega)$ is the product of the coupling strength and spectral density
for that mode.  Note that if the coupling strength is frequency-dependent,
we can absorb that dependence into $g(\omega)$, leaving $\lambda$ as a
frequency-independent parameter.  With this specific choice for $V$, we
can perform the integrals and trace in
(\ref{eq:daviesmaster}--\ref{eq:unitary}).  If we assume that all spacings
between eigenvalues of $H_S$ are distinct, the resulting expression
simplifies considerably, and we find
\bea
  {{\mathrm d}\rho \over {\mathrm d}t} = && -i[H_S,\rho]
    \nonumber\\
    && -\sum_{i,a,b}
    \big[                    N_{ba} |g_{ba}|^2 \<a|\sigma_-^{(i)}|b\>
                                               \<b|\sigma_+^{(i)}|a\>
    \nonumber\\
    && \quad\quad\quad + (N_{ab}+1) |g_{ab}|^2 \<b|\sigma_-^{(i)}|a\>
                                               \<a|\sigma_+^{(i)}|b\> \big]
    \nonumber\\
    && \quad \big\{ (|a\>\<a|\rho) + (\rho|a\>\<a|)
                 - 2 |b\>\<a|\rho|a\>\<b| \big\}
\,,
\label{eq:master}
\eea
where the states $|a\>$ are the time-dependent instantaneous eigenstates
of $H_S$ with energy eigenvalues $\omega_a$,
\be
  N_{ba} = {1 \over \exp[\beta(\omega_b-\omega_a)] - 1}
\ee
is the Bose-Einstein distribution at temperature $1/\beta$, and
\be
  g_{ba} = \cases{
  \begin{array}{ll}
    \lambda g(\omega_b - \omega_a) ~,& \quad \omega_b>\omega_a \\
    0                              ~,& \quad \omega_b \le \omega_a \,.
  \end{array} }
\ee

We simulated the effect of thermal noise by numerically integrating the
master equation (\ref{eq:master}) with a Hamiltonian $H_S$ given by
(\ref{eq:timedep}) and with the initial pure state density matrix
$\rho(0)=|\psi(0)\>\<\psi(0)|$ given by (\ref{eq:start_state}).  For
simplicity, we chose $g(\omega)=1$ for $\omega \ge 0$ and zero otherwise.
Although we would expect that $g(\omega) \to 0$ as $\omega \to \infty$,
for the small systems we are able to simulate it should be a reasonable
approximation to treat $g(\omega)$ as constant and tune the overall
coupling strength using $\lambda^2$.

How should we expect the success probability
$\<\varphi|\rho(T)|\varphi\>$, where $|\varphi\>$ is the ground state of
$H_P$, to depend on the run time $T$ and the temperature?  If the run time
$T$ is sufficiently long, then regardless of its initial state the quantum
computer will come to thermal equilibrium; at the time of the final
readout it will be close to the Gibbs state
\be
  \lim_{T \to \infty}\rho(T) 
   = {e^{-\beta H_P} \over \tr~e^{-\beta H_P}}
   \equiv \rho_P 
\ee
of the problem Hamiltonian $H_P$, and the success probability will be
approximately $\langle \varphi|\rho_P|\varphi\rangle$.  This probability
may be appreciable if the temperature is small compared to the gap between
the ground state and first excited state of $H_P$.  Thus one way to find
the ground state of $H_P$ is to prepare the computer in any initial state,
put it in a cold environment, wait a long time, and measure.  However,
this thermal relaxation method is not an efficient way to solve hard
optimization problems.  Although it may work well on some instances of a
given problem, this method will not work in cases where the computer can
get stuck in local minima from which downward transitions are unlikely.
In such cases, the time for equilibration is expected to be exponentially
large in $n$. 

Consider an instance with a long equilibration time so that cooling alone
is not an efficient way to find the ground state of $H_P$.  It is possible
that the minimum gap $\Delta$ associated with the quantum algorithm is not
small, and the idealized quantum computer, running without decohering
effects, would find the ground state of $H_P$ in a short time.  In this
situation, if we include the coupling of the system to the environment and
we run at a temperature much below $\Delta$, then thermal transitions are
never likely, and the adiabatic algorithm should perform nearly as well as
in the absence of decoherence.  But if the temperature is comparable to
$\Delta$, then the performance may be significantly degraded.

On the other hand, consider an instance for which the equilibration time
is short, so that cooling alone is a good algorithm.  Furthermore, suppose
that the adiabatic algorithm would find the ground state of $H_P$ in a
short time in the absence of decohering effects.  In this case, the
combined effects of cooling and adiabatic evolution will surely find the
ground state of $H_P$ in a short time.  But note that $\Delta$ alone does
not control the success of the algorithm.  Even if $H(t)$ changes too
quickly for the evolution to be truly adiabatic so that a transition
occurs where the gap is smallest, the system may be cooled back into its
ground state at a later time.

\end{multicols}
\begin{figure}
\begin{center}
\begin{tabular}{@{}c@{\hspace{12pt}}c@{}}
\psfig{file=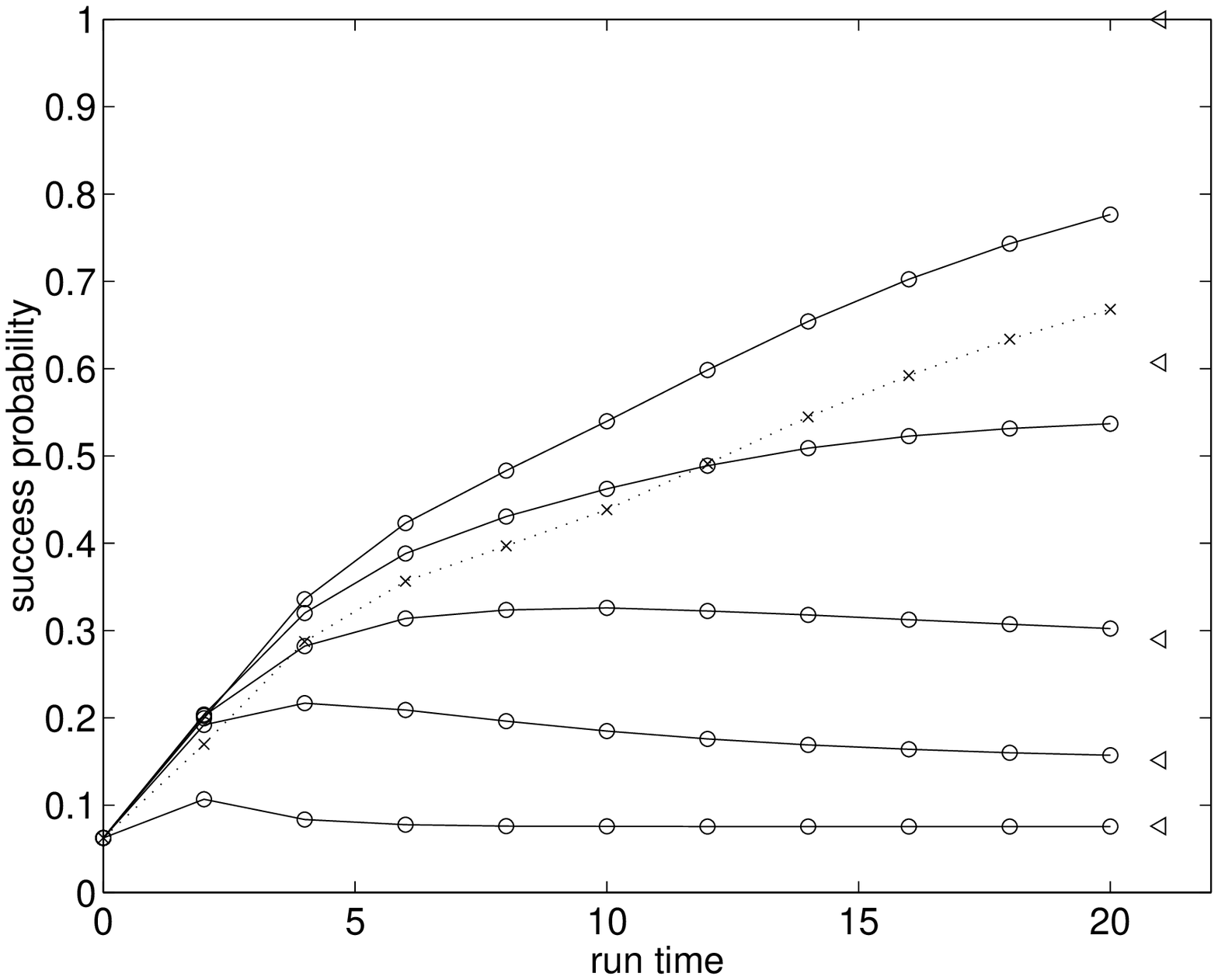,width=3.1in} &
\psfig{file=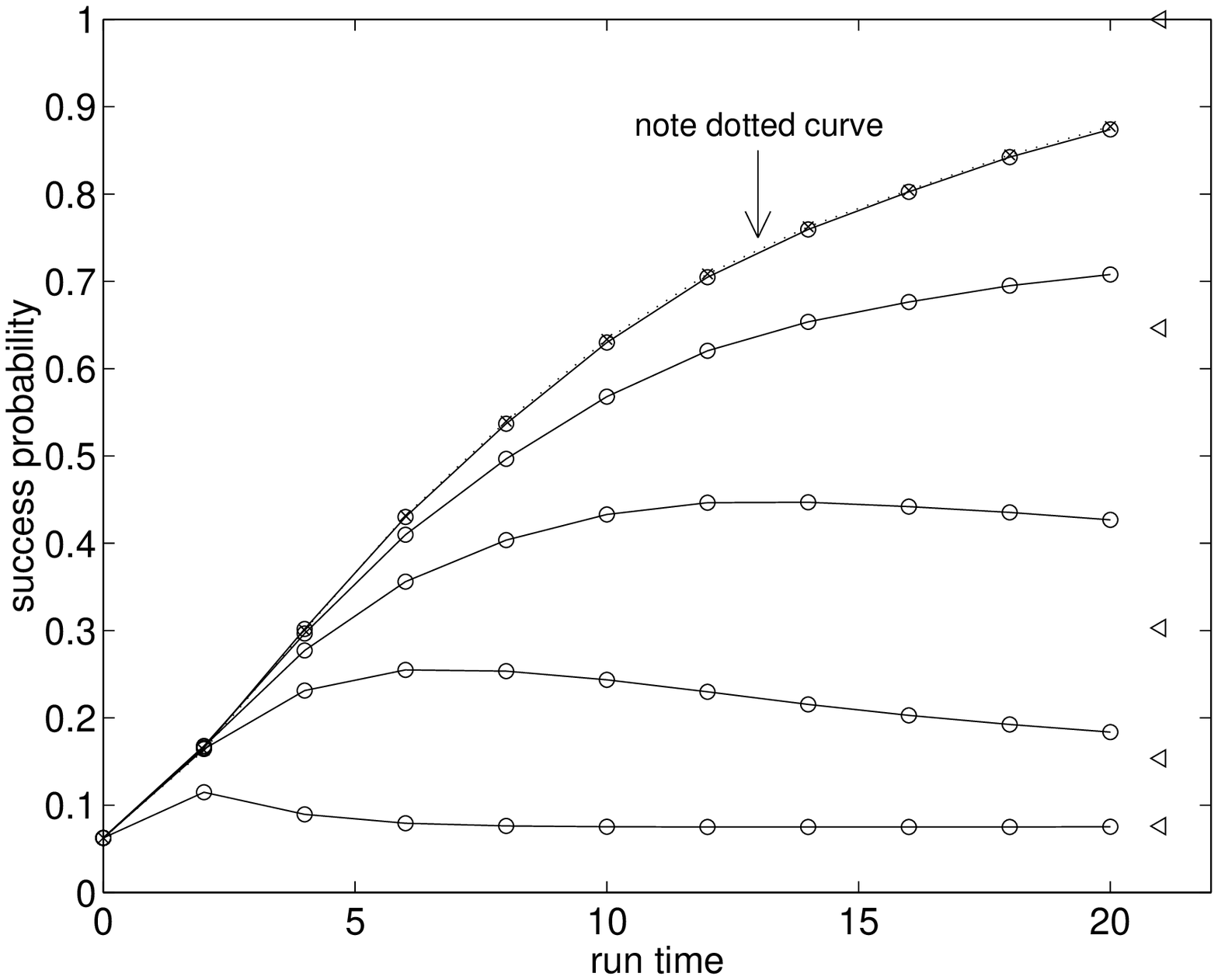,width=3.1in}
\end{tabular}
\end{center}
\caption{The success probability as a function of run time $T$ for two
instances of EC3 with $n=4$ bits.  The instance on the left has a gap of
$\Delta_1 \approx 0.301$ and the instance on the right has a gap of
$\Delta_2 \approx 0.425$.  The dotted line shows the behavior of the
algorithm with no decoherence, i.e., $\lambda^2=0$.  Note that in the
figure on the right, the dotted curve is partially obscured but can be
seen slightly above the topmost solid curve.  The solid lines show the
behavior of the algorithm in the presence of decoherence with
$\lambda^2=0.1$ for five different temperatures.  The triangles at the far
right show the thermal success probabilities $\<\varphi|\rho_P|\varphi\>$
at each of these temperatures.  From top to bottom, the temperatures are
1/10, 1/2, 1, 2, and 10.}
\label{fig:decoherence}
\end{figure}
\begin{multicols}{2}[]

Typical results of the simulation are shown in Fig.~\ref{fig:decoherence}
for two $n=4$ bit instances of EC3 with unique satisfying assignments.
These two instances have minimum gaps of $\Delta_1 \approx 0.301$ and
$\Delta_2 \approx 0.425$.  For each instance, we plot the success
probability as a function of the run time $T$.  With $\lambda^2=0.1$, we
consider five temperatures: 1/10, 1/2, 1, 2, and 10.  We also present the
data with no decoherence ($\lambda^2=0$) for comparison.

Unfortunately, the time required to integrate (\ref{eq:master}) grows very
rapidly with $n$.  Whereas a state vector contains $2^n$ entries, the
density matrix contains $4^n$ entries; and in addition, calculating
${\mathrm d}\rho/{\mathrm d}t$ at each timestep requires evaluating a
double sum over $2^n$ energy eigenstates.  For this reason, we were only
able to consider instances with $n \le 4$.

The results are consistent with our general expectations.  In the absence
of decoherence, the success probability becomes appreciable for
sufficiently long run times.  This probability rises faster for the
problem with a larger gap.  When we add decoherence at high temperature,
the success probability never becomes very large (note the lowest curves
in Fig.~\ref{fig:decoherence}).  As the temperature is decreased to a
value of order one, the presence of decoherence has a less significant
effect on the success probability.  In fact, for sufficiently low
temperatures, the success probability can actually be higher in the
presence of decoherence than when there is no decoherence.  This is
because the primary effect of decoherence at low temperature is to drive
transitions towards the ground state, improving performance.

However, these results do not illustrate a definitive connection between
the minimum gap $\Delta$ and the temperature above which the algorithm no
longer works.  These simple $n=4$ bit instances fall into the second
category discussed above: the equilibration time is short, so cooling
alone is a good algorithm.  In other words, no sharp distinction can be
drawn between the run time required for the adiabatic algorithm to perform
well in the absence of decoherence and the run time required for
equilibration.  Accordingly, the dependence of the success probability on
temperature and run time is similar for the two instances shown in
Fig.~\ref{fig:decoherence}, even though the minimum gaps for these
instances are somewhat different.

\section{Unitary control error}
\label{sec:control_error}

We now consider how the performance of the adiabatic algorithm for EC3 is
affected by adding three different kinds of perturbations to the
Hamiltonian.  Each perturbation we consider is a sum of single-qubit
terms, where each term can be interpreted as a magnetic field pointing in
a random direction.  To simplify our analysis, we assume that the
magnitude of the magnetic field is the same for all qubits, but its
direction varies randomly from qubit to qubit.  The perturbations we
consider are
\bea
  \tilde K_1(s) &=& C_1 s \sum_{i=1}^n \hat m_i \cdot \vec\sigma^{(i)}
  \label{eq:pert1}\,, \\
  \tilde K_2(s) &=& C_2 \sin (\pi s) \sum_{i=1}^n \hat m_i
                                     \cdot \vec\sigma^{(i)}
  \label{eq:pert2}\,, \\
  \tilde K_3(s) &=& {1 \over 2} \sin (C_3 \pi s) \sum_{i=1}^n \hat m_i 
                                                 \cdot \vec\sigma^{(i)}
  \label{eq:pert3}\,,
\eea
which are added to (\ref{eq:full_ham}) and give a time-dependent
Hamiltonian according to (\ref{eq:timedep}).  Each $\hat m_i$ is a
randomly generated real three-component vector with unit length, $C_1$ and
$C_2$ are real numbers, and $C_3$ is a nonnegative integer.

The adiabatic algorithm was simulated by numerically solving the
time-dependent Schr\"odinger equation with initial state $|\psi(0)\>$
given by (\ref{eq:start_state}) and Hamiltonian $\tilde H(t/T) + \tilde
K_j(t/T)$ for a given $j \in \{1,2,3\}$.  As
in~\cite{farhi2,farhi3,farhi4}, we used a fifth-order Runge-Kutta method
with variable step-size, and checked the accuracy by verifying that the
norm of the state was maintained to one part in a thousand.  For a
specified value of $n$, we randomly generated an instance of EC3 with a
unique satisfying assignment.  Then we randomly generated several
different values of the magnetic field directions $\{\hat m_i\}$.  For
each instance of the problem and the magnetic field, the run time was
chosen so that the success probability without the perturbation was
reasonably high.  With this run time fixed, we then determined the success
probability for varying values of the relevant $C_j$.

First, we consider the perturbation $K_1$.  Since it turns on at a
constant rate, this perturbation can be thought of as an error in $H_P$.
Note that with $C_1 \ne 0$, the final Hamiltonian is not simply $H_P$, so
the algorithm will not work exactly even in the adiabatic limit $T \to
\infty$.  This perturbation is potentially dangerous because of the way
its effect scales with the number of bits $n$.  Indeed, consider the case
where $H_P$ can be separated into a sum of Hamiltonians acting separately
on each qubit.  If adding $K_1$ reduces the overlap of the ground state
$|\varphi\>$ of $H_P$ with the perturbed ground state $|\varphi'\>$ by
some fixed value $\epsilon$ for each of the $n$ qubits, then the total
overlap is $(1-\epsilon)^n$, which is exponentially small in the number of
bits.  Thus the algorithm clearly fails in this factorized case.  In
general, if the magnitude of $K_1$ is independent of $n$, then we expect
the algorithm to fail.  However, if the magnitude of $K_1$ falls as $1/n$
or faster, then the shift of the ground state may be small enough (as it
would be in the factorized case) that the algorithm is not significantly
affected.  Note that for any $n$ there is some value of $C_1$ that is
small

\end{multicols}
\begin{figure}
\begin{center}
\begin{tabular}{@{}c@{\hspace{12pt}}c@{}}
  $n=7$ & $n=10$ \\
  \psfig{file=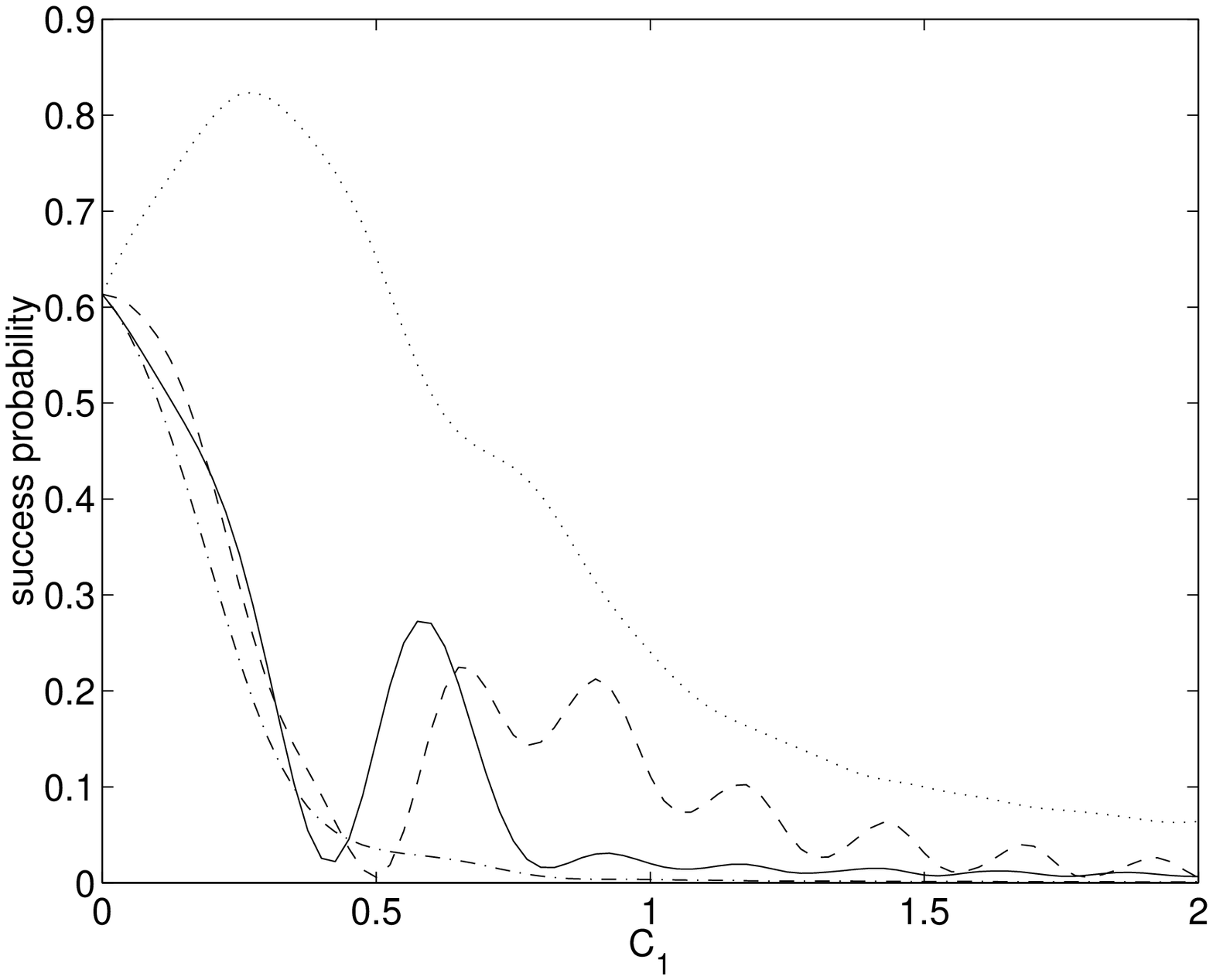,width=3in} &
  \psfig{file=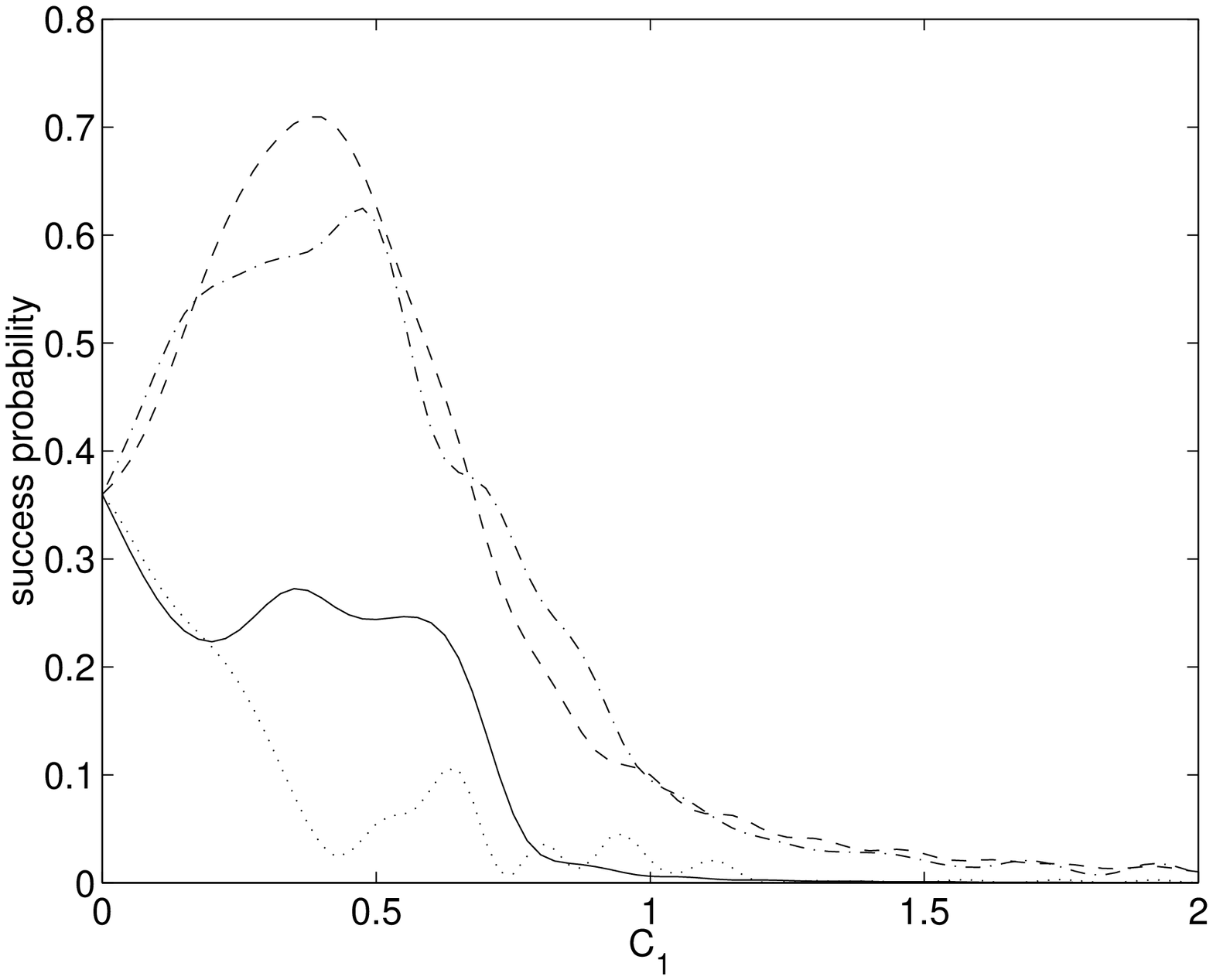,width=3in} \\
  \psfig{file=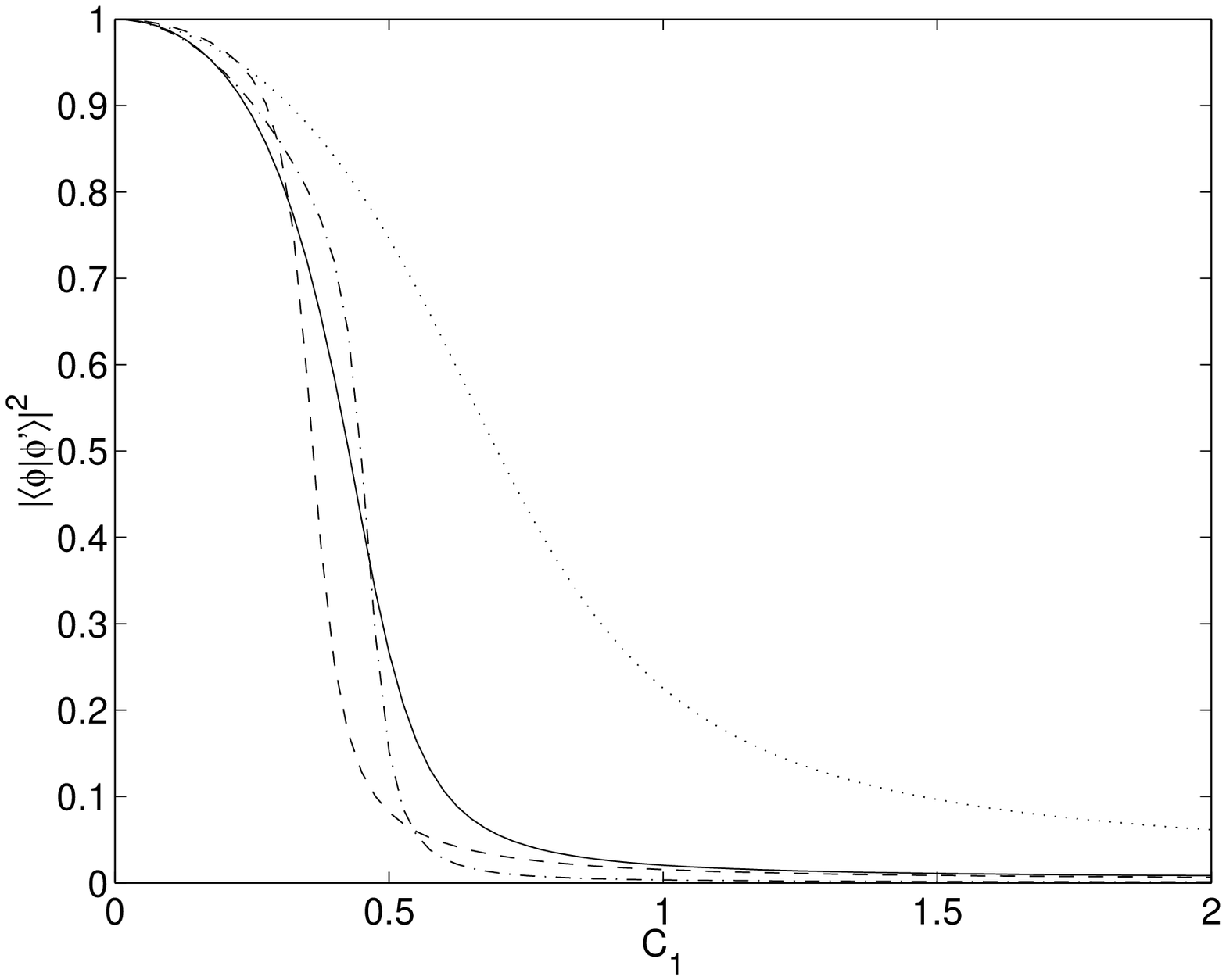,width=3in} &
  \psfig{file=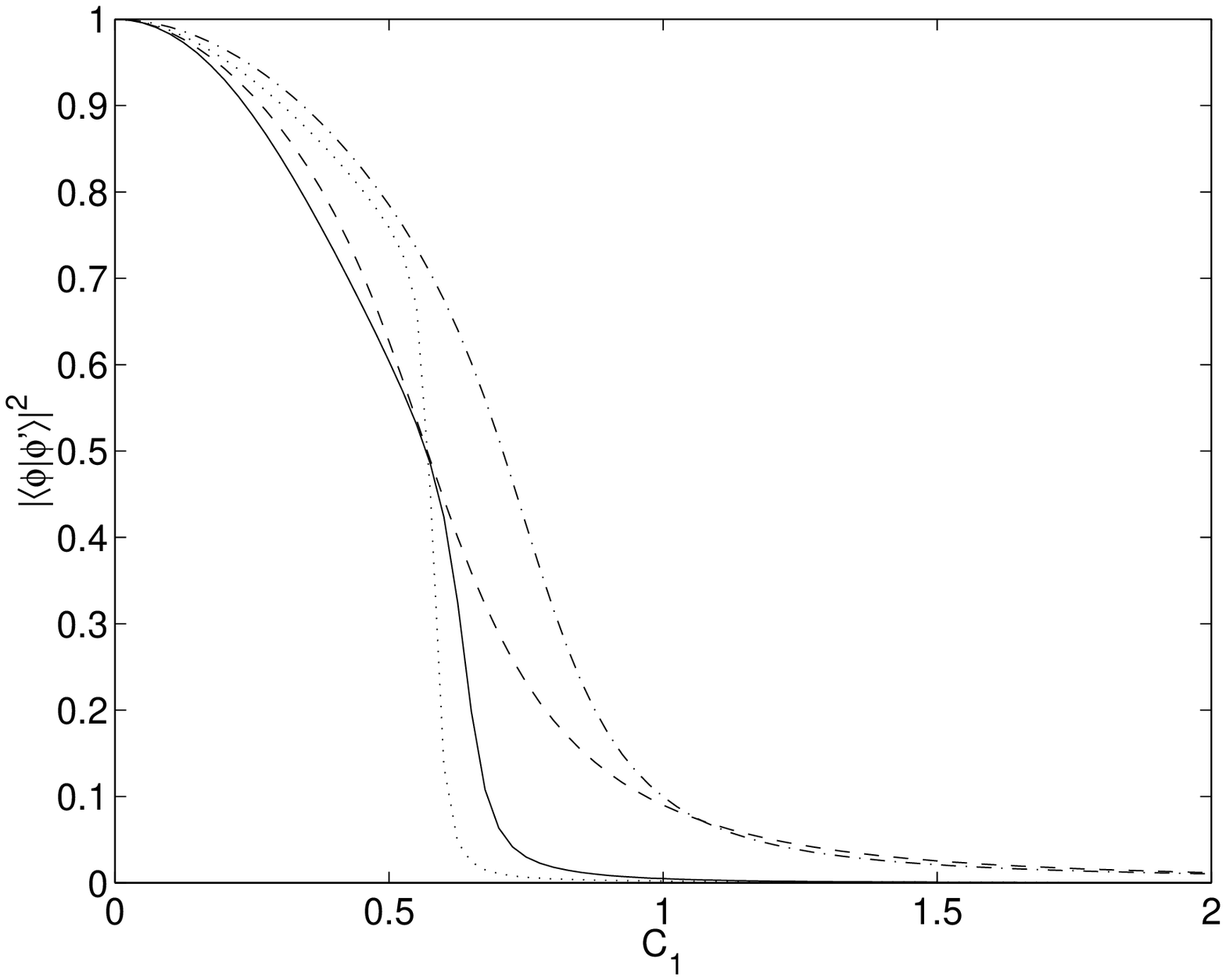,width=3in}
\end{tabular}
\end{center}
\caption{(Top) The success probability of the adiabatic algorithm for two
randomly generated instances of EC3 with $n=7$ bits (left) and $n=10$ bits
(right) under the perturbation $K_1$ defined by (\ref{eq:pert1}) for four
different sets of magnetic field directions.   For each $n$, the run time
is the same for each random perturbation.  (Bottom) The corresponding
overlaps $|\<\varphi|\varphi'\>|^2$ of the ground state $|\varphi\>$ of
$H_P$ with the perturbed ground state $|\varphi'\>$ at $s=1$.}
\label{fig:C1}
\end{figure}
\begin{multicols}{2}[]

\end{multicols}
\begin{figure}
\begin{center}
\begin{tabular}{@{}c@{\hspace{12pt}}c@{}}
  $n=7$ & $n=10$ \\
  \psfig{file=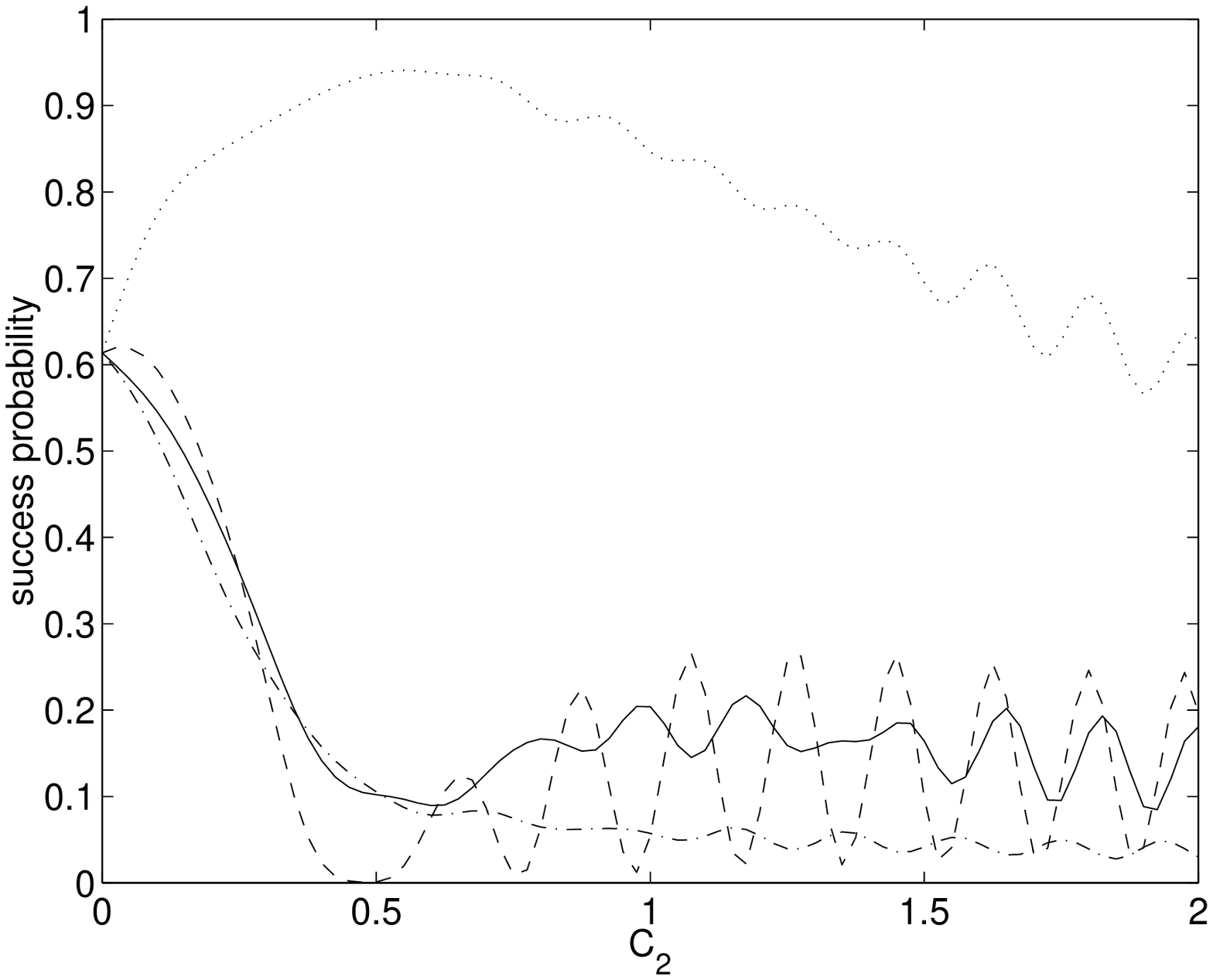,width=3in} &
  \psfig{file=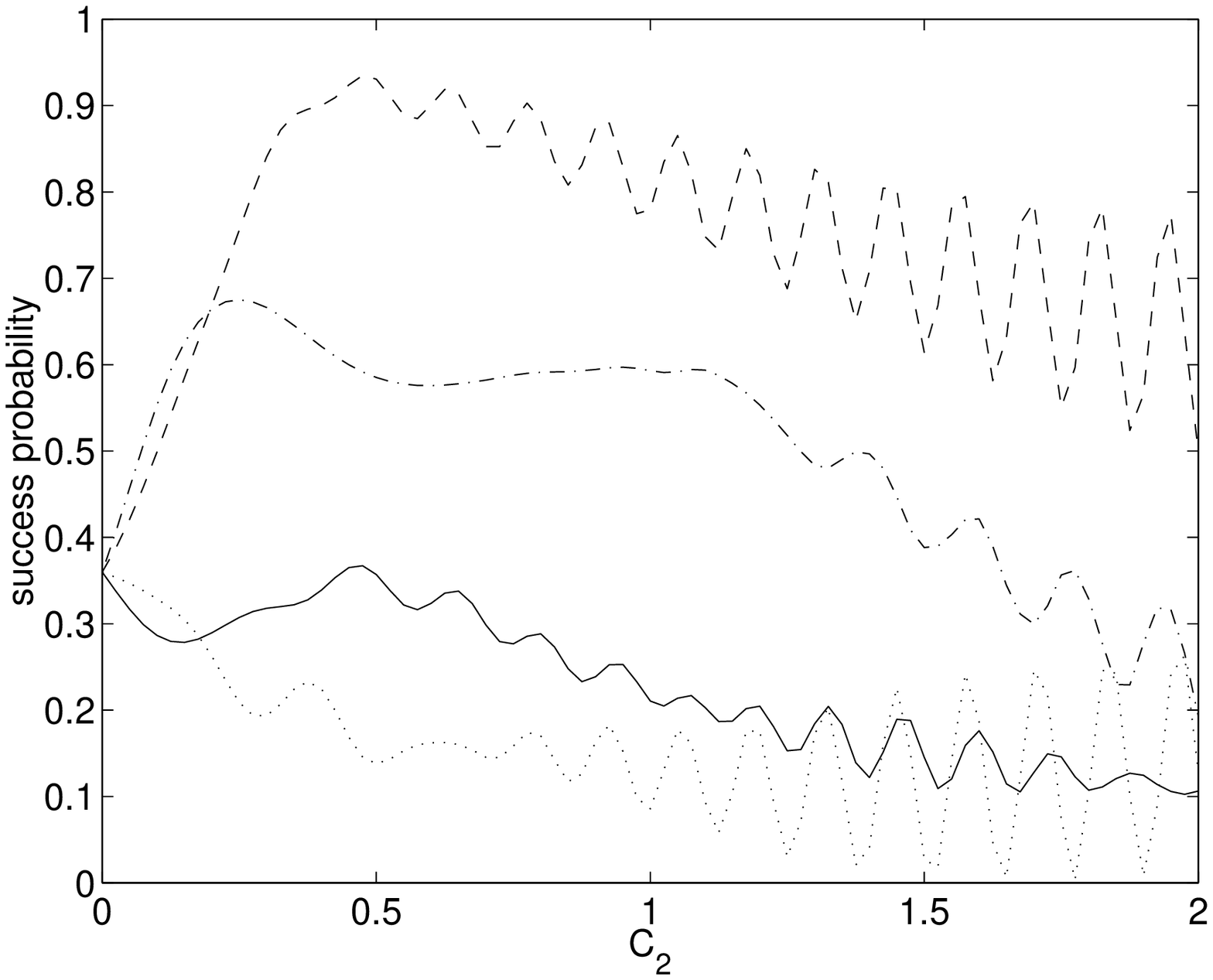,width=3in} \\
  \psfig{file=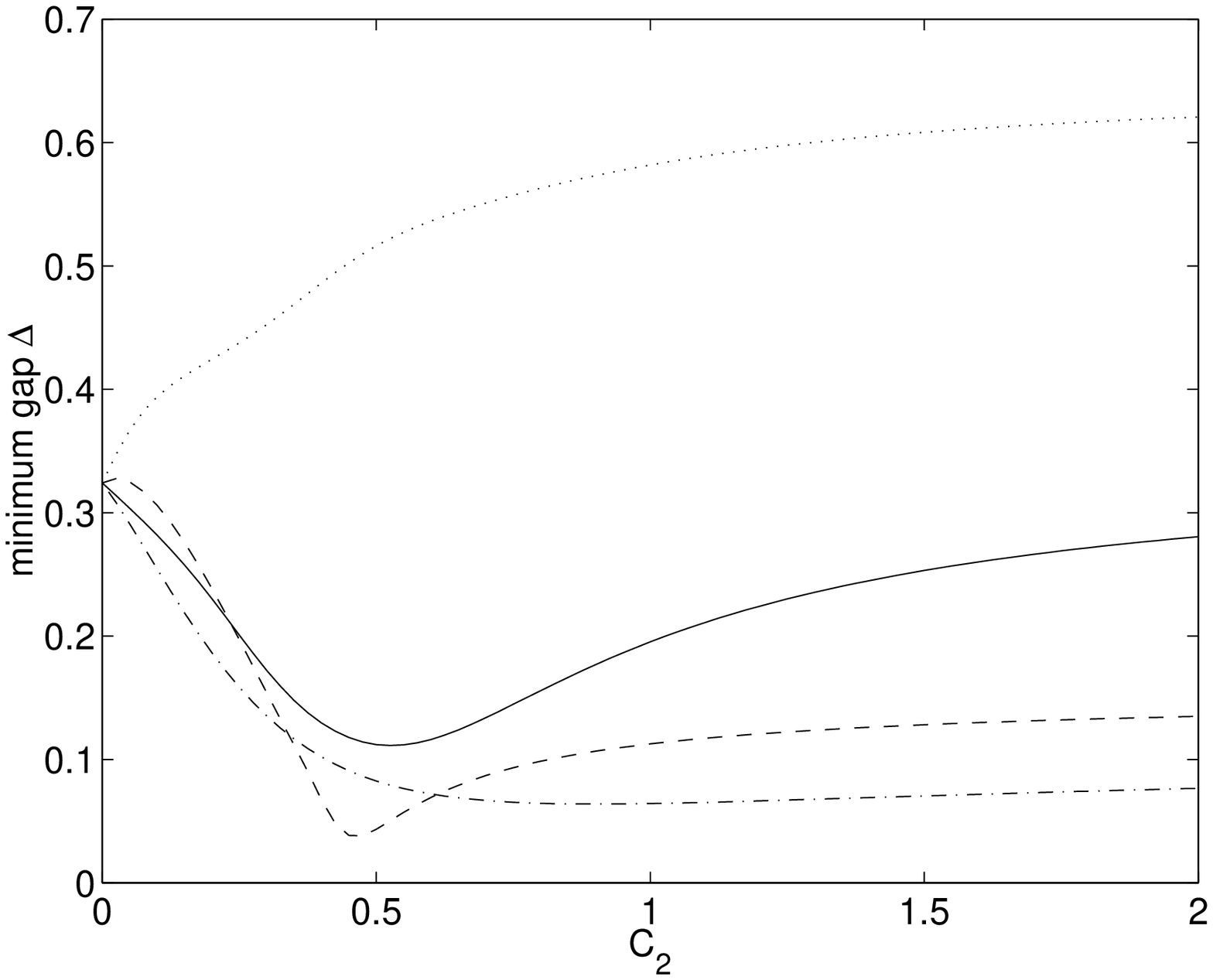,width=3in} &
  \psfig{file=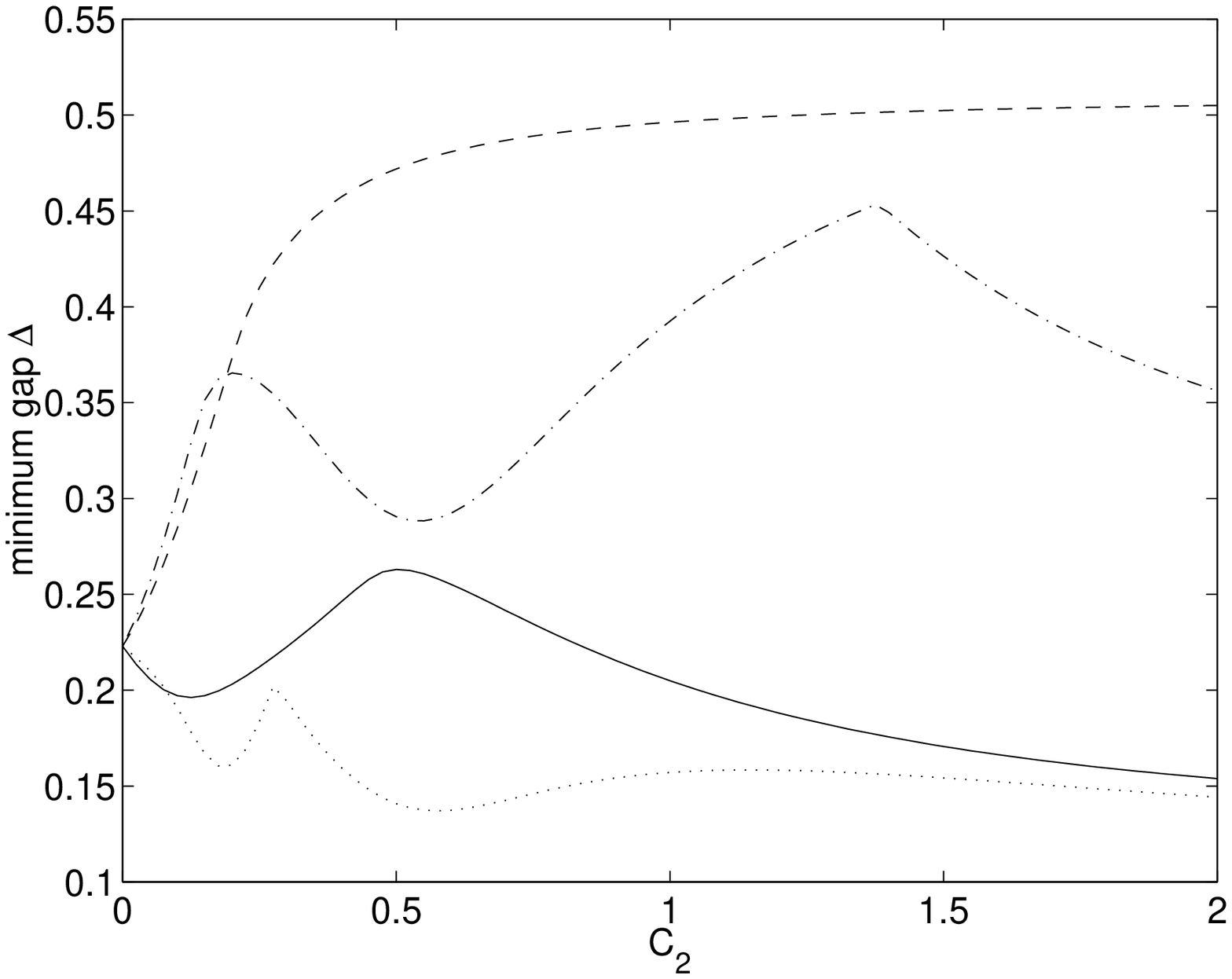,width=3in}
\end{tabular}
\end{center}
\caption{(Top) The success probability of the adiabatic algorithm for the
same instances used in Fig.~\ref{fig:C1} under the perturbation $K_2$
defined by (\ref{eq:pert2}).  The four different magnetic field directions
for each instance are also the same as in Fig.~\ref{fig:C1}.  (Bottom) The
minimum gap $\Delta$ in the perturbed problem.}
\label{fig:C2}
\end{figure}
\begin{multicols}{2}[]

\noindent 
enough that the disadvantage of reduced overlap with the ground state of
$H_P$ may be overcome if the perturbation happens to increase the minimum
gap $\Delta$.  For this reason, we expect to sometimes see an increase in
success probability for small $C_1$ that goes away as $C_1$ is increased.

The effect of the perturbation $K_1$ is shown in Fig.~\ref{fig:C1} for
$n=7$ and $n=10$ bit instances of EC3, with four different randomly
generated sets of magnetic field directions for each instance.  The run
time is chosen such that for $C_1=0$, the success probability is around
$1/2$.  The top plots show that for small $C_1$, the success probability
is not strongly suppressed; in fact, in some cases it is significantly
enhanced.  For large enough $C_1$, the success probability is heavily
suppressed.  The bottom plots show the overlap $|\<\varphi|\varphi'\>|^2$
between the ground state of $H_P$ and the actual ground state in the
presence of the perturbation.  As we
expect, the suppression of the success probability is correlated with the
amount of overlap.  We also studied a similar perturbation in which $s$ is
replaced by $1-s$, which can be thought of as an error in $H_B$.
Unsurprisingly, the results were qualitatively similar.

Next, we consider the low frequency perturbation $K_2$.  The period of
oscillation is chosen such that the perturbation vanishes at $t=0$ and
$t=T$, so the perturbation does not affect the algorithm in the adiabatic
limit.  Since the success probability is quite sensitive to the value of
the minimum gap $\Delta$, and it is not {\em a priori} obvious whether a
perturbation will increase or decrease $\Delta$, we can guess that turning
on a nonzero value of $C_2$ can either increase the success probability or
decrease it.

Fig.~\ref{fig:C2} shows the effect of the perturbation $K_2$, using the
same instances, magnetic field directions, and run times as in
Fig.~\ref{fig:C1}.  The top plots show the success probability as a
function of $C_2$.  As in the case of $K_1$, some perturbations can raise
the success probability and some suppress it.  Perhaps unsurprisingly, a
particular set of magnetic field directions that can raise the success
probability under $K_1$ is also likely to help when $K_2$ is applied.  But
unlike $K_1$, $K_2$ can improve the success probability even with $C_2
\simeq 2$, where the size of the perturbation is comparable to the size of
the unperturbed Hamiltonian.  The bottom plots show the minimum gap
$\Delta$ when the perturbation is added.  Note that there is a strong
correlation between the success probability and $\Delta$.

For both perturbations $K_1$ and $K_2$, similar results have been observed
(with fewer data points) for instances with as many as $n=14$ bits.
Figs.~\ref{fig:C1} and \ref{fig:C2} present typical data.  For example,
for a given instance, typically one or two out of four sets of randomly
chosen magnetic field directions led to an improvement in the success
probability for some values of $C_1$ and $C_2$, compared to the
unperturbed case.

Finally, we consider the perturbation $K_3$, in which the magnitude of the
oscillating component is fixed, but we may vary its frequency by varying
$C_3$.  As for $K_2$, the frequency is chosen so that the perturbation
vanishes at $t=0$ and $t=T$.  We expect that for $C_3$ of order one, the
perturbation will be likely to excite a transition, and that the success
probability will be small.  But since both $H_B$ and $H_P$ have a maximum
eigenvalue of order $n$, we can anticipate that for
\be
  C_3 \gg {nT \over \pi}
\,,
\label{eq:high_condition}
\ee
the perturbation will be far from any resonance.  Then the probability
that the perturbation drives a transition will be low, and the success
probability should be comparable to the case where the perturbation
vanishes. 

\begin{figure}
\begin{center}
\begin{tabular}{@{}c@{\hspace{12pt}}c@{}}
  $n=8$ & $n=10$ \\
  \psfig{file=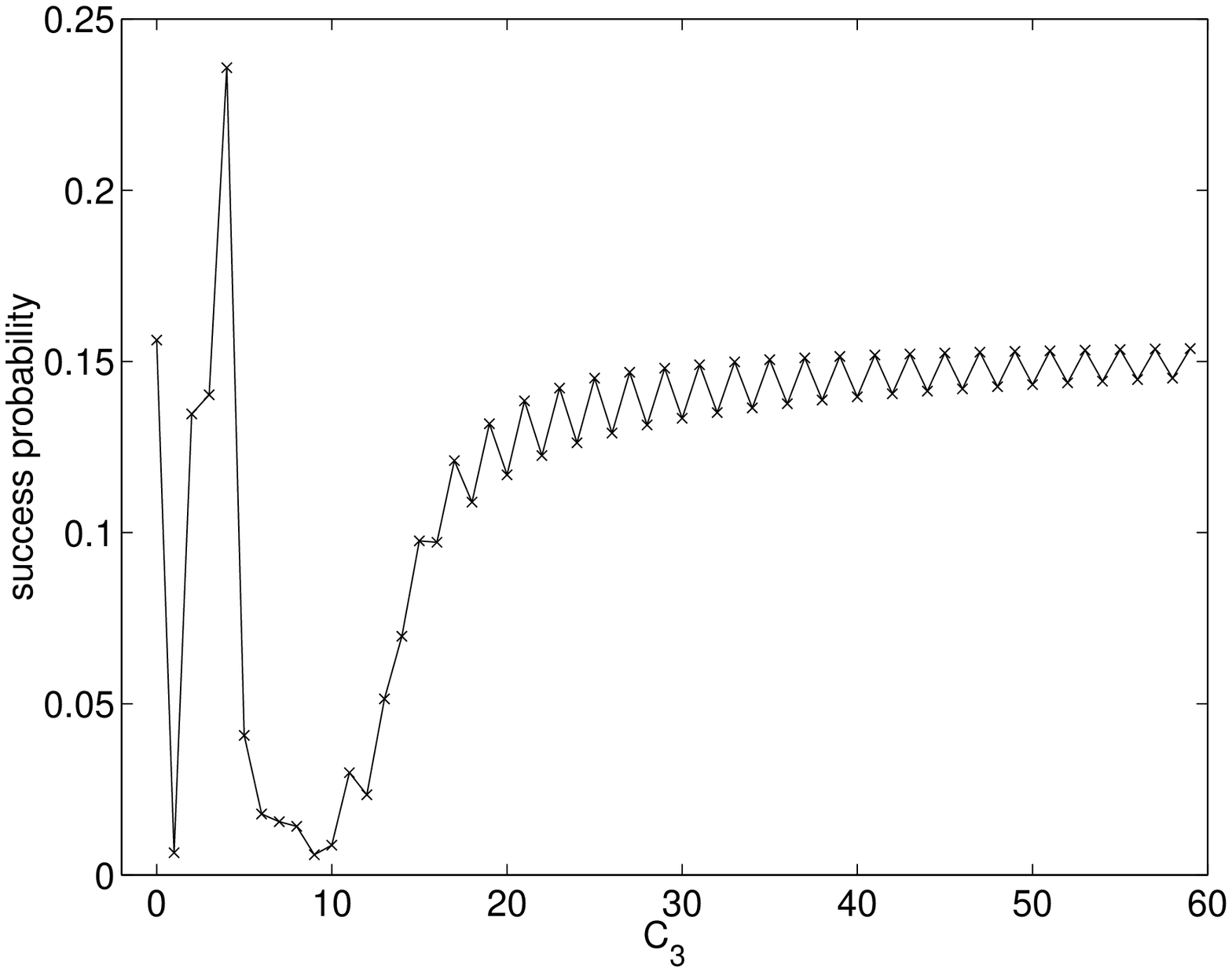,width=3in} &
  \psfig{file=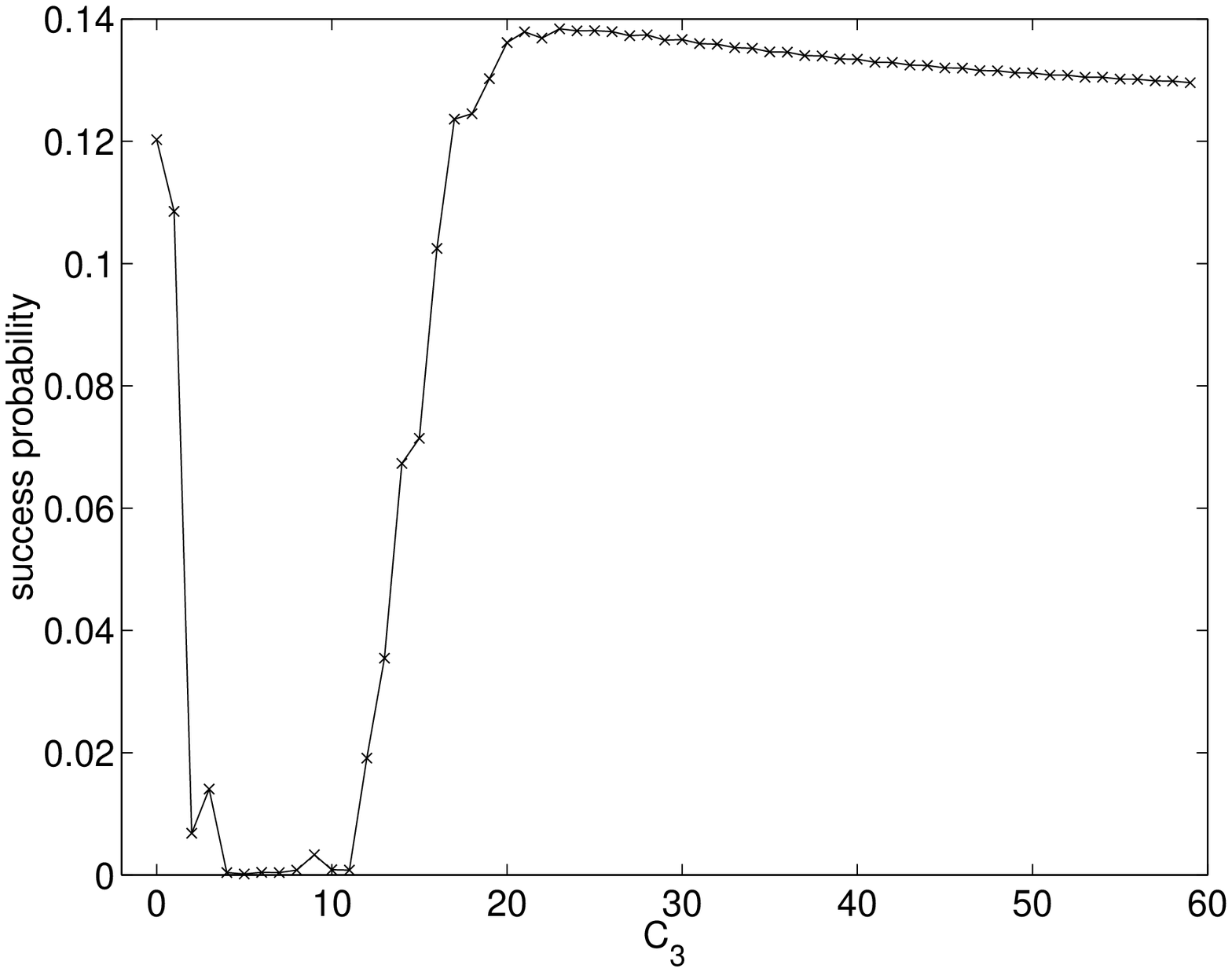,width=3in} \\
  \psfig{file=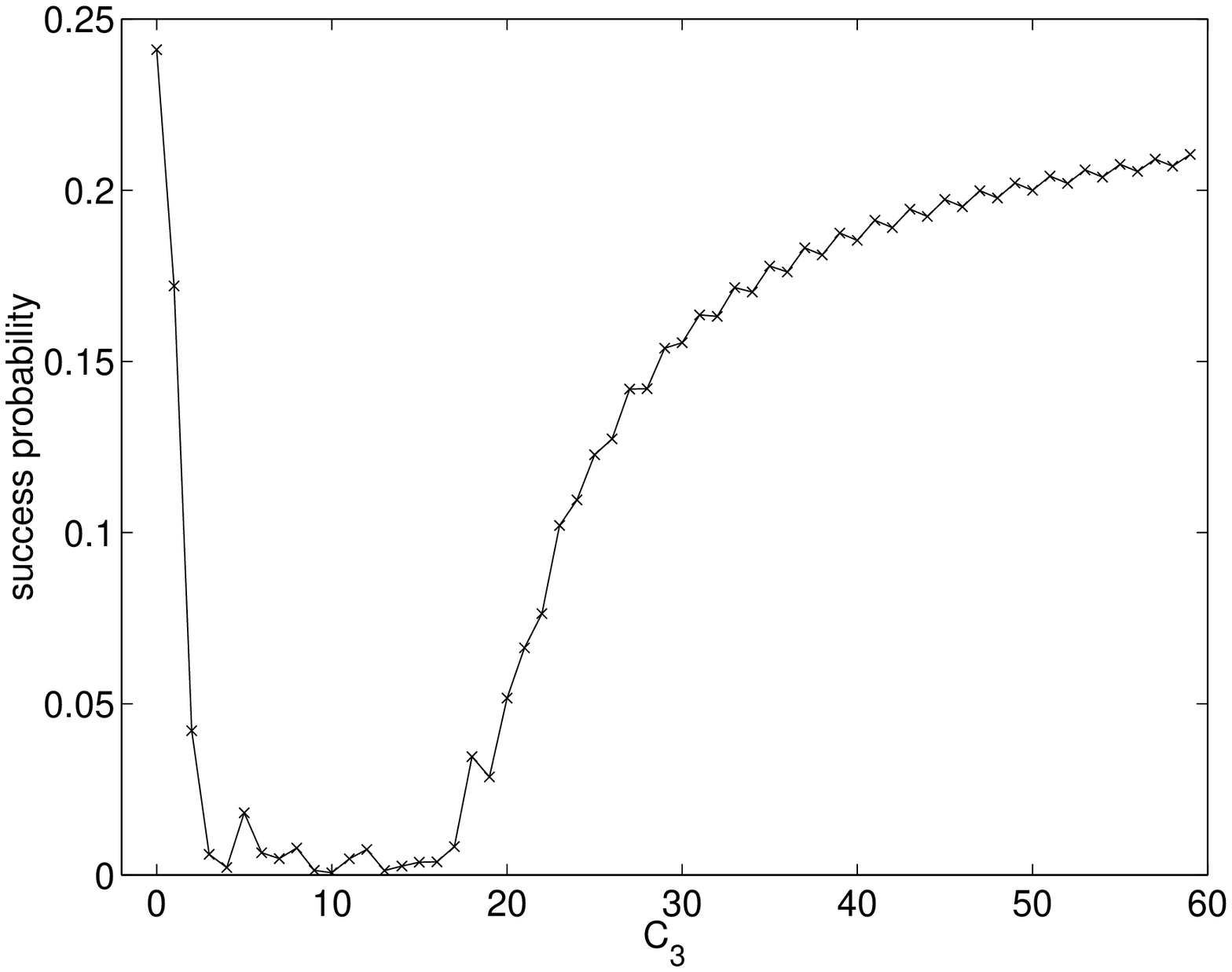,width=3in} &
  \psfig{file=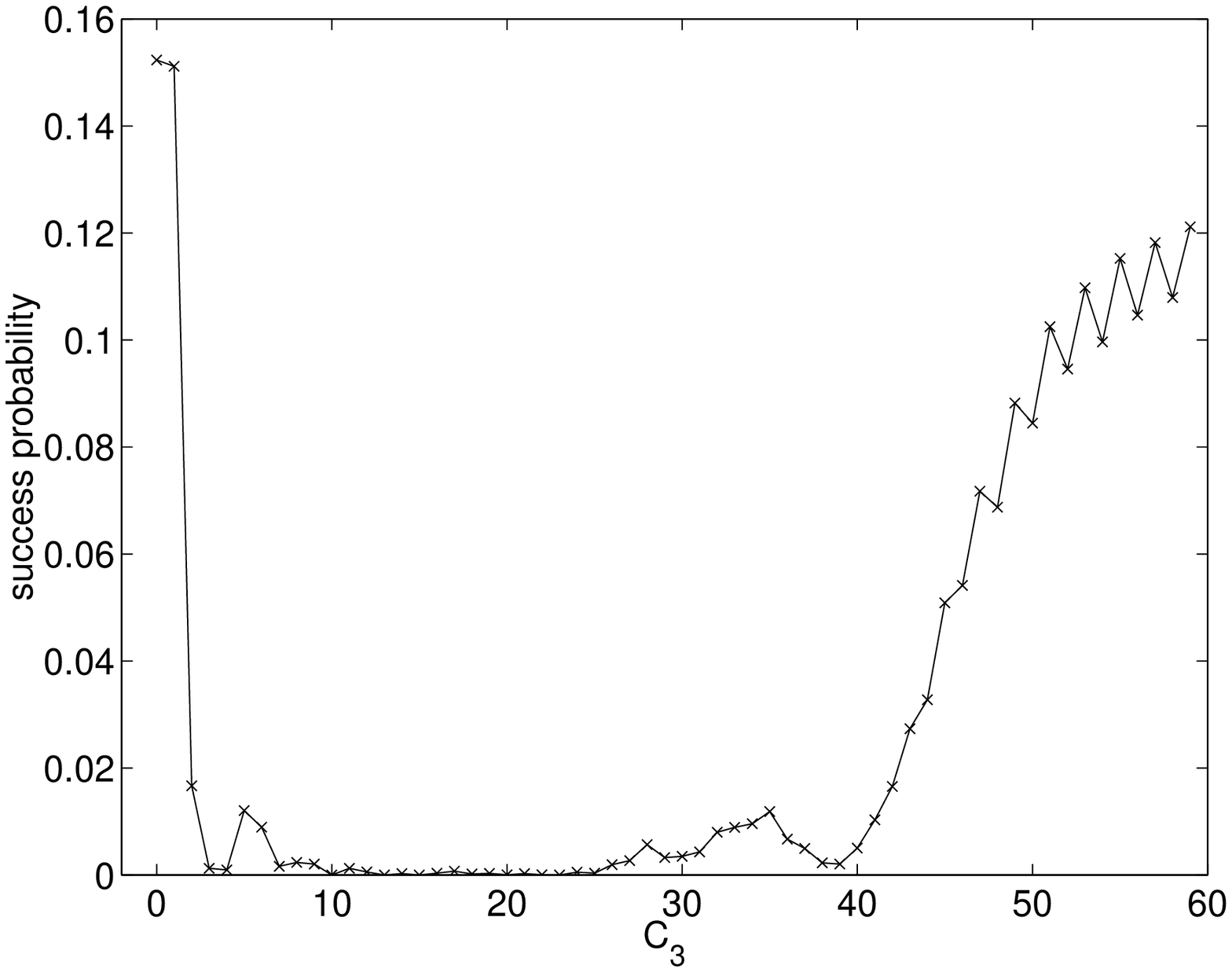,width=3in}
\end{tabular}
\end{center}
\caption{The success probability as a function of the frequency $C_3$ of
the perturbation $K_3$ defined in (\ref{eq:pert3}).  The data in each plot
were obtained for a randomly generated instance of EC3 with randomly
generated magnetic field directions.  The data in the left column are for
two instances with $n=8$ bits, and the data in the right column are for
two instances with $n=10$ bits.  For the top row, the run time is chosen
so that the success probability is around $1/8$ for $C_3=0$, and for the
bottom row, the run time is twice as long.  The leftmost points in each
plot correspond to $C_3=0$, so the perturbation is absent for all $t$.
$C_3$ takes integer values, so the lines are included only to guide the
eye.}
\label{fig:C3}
\end{figure}

Some representative plots of the dependence of the success probability on
$C_3$ are shown in Fig.~\ref{fig:C3}.  Each plot corresponds to a
particular randomly generated instance of EC3 (with either $n=8$ bits or
$n=10$ bits) and a randomly generated set of magnetic field directions.
In the top row of plots, the run time is chosen so that the success
probability is around $1/8$ with the perturbation absent (i.e., $C_3=0$).
In the bottom row, the run time is

\end{multicols}
\begin{multicols}{2}[]

\noindent
doubled.  All of the data exhibit the expected qualitative trend.  The
leftmost point corresponds to $C_3=0$.  For the smallest values of
$C_3>0$, the success probability may not be too badly damaged; for
somewhat larger values of $C_3$ it is heavily suppressed; and for
sufficiently large $C_3$ it recovers to a value near the success
probability in the absence of the perturbation.  The value of $nT/\pi$ is
around 19 and 39 for the upper and lower $n=8$ plots and is around 38 and
76 for the upper and lower $n=10$ plots, so the estimate
(\ref{eq:high_condition}) turns out to be reasonable.

Another conspicuous feature of the plots in Fig.~\ref{fig:C3} is that the
success probability tends to oscillate between even and odd values of
$C_3$, though whether even or odd values are favored varies from case to
case.  This occurs because the perturbation's time average vanishes for
$C_3$ even, so that its integrated effect is weaker than for $C_3$ odd.
Since a small perturbation might either help or hurt, the success
probability is slightly enhanced for odd $C_3$ in some cases, and is
slightly suppressed in other cases.

\section{Conclusions}
\label{sec:conclusions}

We have conducted numerical simulations to investigate the fault tolerance
of adiabatic quantum computation, and our results are consistent with the
claim that this algorithm is robust against decoherence and certain kinds of
random unitary perturbations.  Thus, if a physical system could be
engineered with interactions reasonably well described by a Hamiltonian that
smoothly interpolates from an initial $H_B$ to a final $H_P$ corresponding
to an interesting combinatorial search problem, and if the gap remains large
throughout the interpolation, that system might be a powerful computational
device.

Although we have viewed unitary perturbations as noise, the fact that they
sometimes raise the success probability suggests a possible way to speed
up the adiabatic algorithm.  The algorithm finds the ground state of $H_P$
by starting the system in the ground state of $H_B$.  The quantum state
evolves as the system Hamiltonian smoothly interpolates from $H_B$ to
$H_P$.  However, there are many possible choices for $H_B$ and many smooth
paths from a given $H_B$ to $H_P$.  The choices (\ref{eq:hb}) and
(\ref{eq:full_ham}) are convenient but arbitrary, so choosing an alternate
route to $H_P$ might speed up the algorithm.  An example of this is seen
in~\cite{RC01}, where it is shown that optimizing the time-dependent
coefficients of $H_B$ and $H_P$ allows the adiabatic algorithm to achieve
a square root speedup for an unordered search problem.  More generally,
the interpolating Hamiltonian might involve terms which have nothing to do
with $H_B$ or $H_P$, but which increase $\Delta$ and therefore improve
performance.  For example, the perturbation $K_2$ sometimes increases the
success probability, as seen in Fig.~\ref{fig:C2}.  Rather than being
thought of as a source of error, such a perturbation could be applied
intentionally and might sometimes enhance the effectiveness of the
adiabatic algorithm.

\acknowledgments

We thank Todd Brun, Evan Fortunato, Jeffrey Goldstone, Sam Gutmann, Jeff
Kimble, Alesha Kitaev, and Seth Lloyd for helpful discussions.  AMC
gratefully acknowledges the support of the Fannie and John Hertz
Foundation.  This work has been supported in part by the Department of
Energy under Grant No.\ DE-FG03-92-ER40701 and Grant No.\
DE-FC02-94-ER40818, by the National Science Foundation under Grant No.\
EIA-0086038, by the Caltech MURI Center for Quantum Networks under ARO
Grant No.\ DAAD19-00-1-0374, by the National Security Agency (NSA) and
Advanced Research and Development Activity (ARDA) under Army Research
Office (ARO) contract number DAAD19-01-1-0656, and by an IBM Faculty
Partnership Award.


\end{multicols}
\end{document}